\documentclass[prl,aps,a4paper,twocolumn,superscriptaddress,nofootinbib]{revtex4-1}

\usepackage[utf8x]{inputenc}
\usepackage{amssymb}
\usepackage{amsmath}
\usepackage{graphicx}
\usepackage{bbm}
\usepackage{psfrag}
\usepackage{latexsym}
\usepackage[usenames, dvipsnames]{color}
\usepackage{xcolor}	
\usepackage{hyperref}			
\hypersetup{
    colorlinks,
    linkcolor={blue!50!black},
    citecolor={blue!50!black},
    urlcolor={blue!80!black}
}

\newcommand{\be}{\begin{equation}}
\newcommand{\ee}{\end{equation}}
\newcommand{\bea}{\begin{eqnarray}}
\newcommand{\eea}{\end{eqnarray}}
\newcommand{\ba}{\begin{eqnarray}}
\newcommand{\ea}{\end{eqnarray}}

\newcommand{\beq}{\begin{equation}}
\newcommand{\eeq}{\end{equation}}
\newcommand{\beqa}{\begin{eqnarray}}
\newcommand{\eeqa}{\end{eqnarray}}
\newcommand{\beqar}{\begin{eqnarray*}}
\newcommand{\eeqar}{\end{eqnarray*}}

\newcommand{\eq}{\begin{equation}}
\newcommand{\eqx}{\end{equation}}
\newcommand{\eqn}{\begin{eqnarray}}
\newcommand{\eqnx}{\end{eqnarray}}

\usepackage[normalem]{ulem}

\begin{document}

\author{Pablo A. Cano}\email{pablo.cano@uam.es}

\affiliation{\it Perimeter Institute for Theoretical Physics, Waterloo, Ontario N2L 2Y5, Canada}

\affiliation{\it Instituto de F\'isica Te\'orica UAM/CSIC, C/ Nicol\'as Cabrera, 13-15, C.U. Cantoblanco, 28049 Madrid, Spain}

\author{Robie A. Hennigar}\email{rhennigar@uwaterloo.ca}

\affiliation{\it Department of Physics $\&$ Astronomy, University of Waterloo, Waterloo, ON N2L 3G1, Canada}

\author{Hugo Marrochio}\email{hmarrochio@perimeterinstitute.ca}

\affiliation{\it Perimeter Institute for Theoretical Physics, Waterloo, Ontario N2L 2Y5, Canada}

\affiliation{\it Department of Physics $\&$ Astronomy, University of Waterloo, Waterloo, ON N2L 3G1, Canada}

\title{Complexity Growth Rate in Lovelock Gravity}

\begin{abstract}
Using the ``Complexity = Action'' framework we compute the late time growth of complexity for charged black holes in Lovelock gravity. Our calculation is facilitated by the fact that the null boundaries of the Wheeler-DeWitt patch do not contribute at late times and essential contributions coming from the joints are now understood \cite{CanoJoints}. The late time growth rate reduces to a difference of internal energies associated with the inner and outer horizons, and in the limit where the mass is much larger than the charge, we reproduce the celebrated result of $2M/\pi$ with corrections proportional to the highest Lovelock coupling in even (boundary) dimensions. We find in some cases a minimum mass below which complexity remains effectively constant, even if the black hole contains a non-degenerate horizon. 

\end{abstract}

\maketitle

The Anti de Sitter/Conformal Field Theory (AdS/CFT) duality \cite{Maldacena:1997re} has brought surprising insight to the nature of quantum gravity. The entanglement properties of the CFT suggest an emergence of the bulk geometry \cite{VanRaamsdonk:2010pw}, with the Ryu-Takayanagi prescription for the entanglement entropy (EE) being the most explored example \cite{Ryu:2006ef, Nishioka:2009un, Hubeny:2007xt}. From the seminal works of Bekenstein and Hawking \cite{Bekenstein:1973ur, Bardeen:1973gs, Bekenstein:1974ax, Hawking:1974sw}, the thermodynamic properties of black hole geometries have raised intriguing questions about the possible microscopic structure of black holes.  Understanding how these quantities are encoded in the dual CFT is presently an active area of study \cite{Maldacena:2001kr, Casini:2011kv, Lashkari:2013koa, Hartman:2013qma, Faulkner:2013ana, Faulkner:2013ica, Maldacena:2013xja, Swingle:2014uza, Almheiri:2014lwa, Jacobson:2015hqa, Pastawski:2015qua, Czech:2015qta, deBoer:2015kda}.
 
\begin{figure}
\centering
\includegraphics[width=0.45\textwidth]{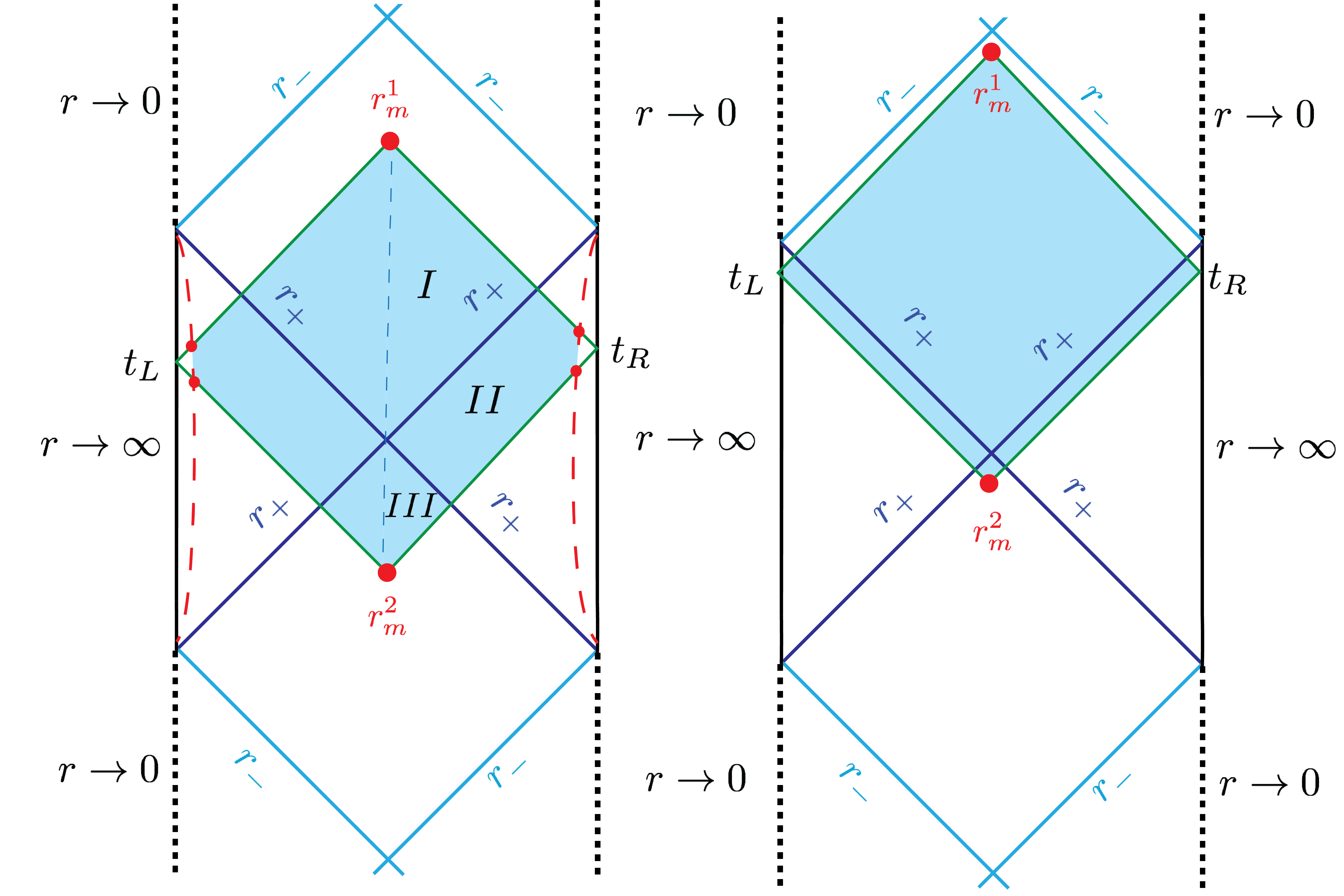}
\caption{The causal structure for a charged AdS black hole, with outer and inner horizons. The blue shaded region denotes the WDW patch, anchored at the boundary times $t_L=t_R=t/2$. At late times, the null boundaries of the WDW patch approach the inner and outer horizons.}
\label{PenroseCharged}
\end{figure}

Recently, it has been observed that entanglement entropy may not be the appropriate observable to probe the interior degrees of freedom of black holes \cite{Susskind:2014moa}. Based on intuition brought by tensor network models in holography \cite{MERA, Swingle:2009bg}, it was proposed that the complexity of the CFT state should encode information about the semi-classical geometry in the interior \cite{Susskind:2014rva, Harlow:2013tf}, motivated originally by the AMPS paradox \cite{Almheiri:2012rt}. Roughly speaking, complexity measures how hard it is to construct certain states in the theory from simple unentangled states and a few universal gates.\footnote{See \cite{Stanford:2014jda, Susskind:2014jwa, Brown:2017jil, Jefferson:2017sdb, Chapman:2017rqy, Caputa:2017urj, Caputa:2017yrh, Yang:2017nfn, Khan:2018rzm} for consideration on properties of complexity in the boundary theory.} There are two related proposals conjectured to capture the complexity of the ground states of CFT in holography: in this letter we focus on the ``complexity$=$action'' (CA) proposal \cite{Brown:2015bva, Brown:2015lvg},\footnote{For exploration on ``complexity$=$volume'', see~\cite{Susskind:2014rva, Stanford:2014jda, Susskind:2014jwa,TimeDep, Couch:2016exn}. } as it permits a conceptually straightforward generalization to include gravitational higher curvature corrections. The CA proposal states that the complexity of the state is given by the Lorentzian action evaluated on the Wheeler-DeWitt (WDW) patch,
\begin{equation}
\mathcal{C}_{A} = \frac{I_{\rm WDW}}{\pi} \, ,
\end{equation}
which is the union of all spacelike hypersurfaces anchored at boundary times $t_L$ and $t_R$, as shown in figure~\ref{PenroseCharged}.

Complexity is conjectured to continue increasing long after local thermal equilibrium is reached. The late time rate of change is approximately $2 M/\pi$ in CA for a large class of uncharged black holes in Einstein Gravity (EG)~\cite{Brown:2015bva, Brown:2015lvg, NullBoundaries}, suggesting a connection to Lloyd's bound on the rate of computation \cite{Lloyd},\footnote{Note however that it has been pointed out that the Lloyd bound can be violated in certain circumstances~\cite{TimeDep, Couch:2017yil,Swingle:2017zcd, Alishahiha:2018tep,An:2018xhv}.} and that black holes are the fastest scramblers in nature \cite{Sekino:2008he, Brown:2015bva, Brown:2015lvg}.

While the properties of complexity in EG are now well understood in many situations, relatively little is known about its behavior in higher-derivative gravity \cite{Cai:2016xho, Alishahiha:2017hwg}.  In the context of AdS/CFT,  higher-order curvature corrections in the bulk are dual to finite $N$ and finite coupling effects in the CFT \cite{Grisaru:1986px,Gross:1986iv,Gubser:1998nz,Buchel:2004di}. The most relevant aspect of these corrections is that they allow us to explore more general holographic CFTs than those defined by Einstein gravity \cite{Camanho:2009vw,deBoer:2009pn,Buchel:2008vz,Hofman:2008ar,Hofman:2009ug,Nojiri:1999mh,Blau:1999vz,Buchel:2009sk,Myers:2010jv,Bueno:2018xqc}.  
In this way, higher-order gravities allow us to identify universal relations valid for arbitrary theories, including for example general results about the EE \cite{Myers:2010tj,Myers:2010xs,Mezei:2014zla,Bueno1,Bueno2}. Hence, one may wonder if, in the case of complexity, higher-curvature gravities could help us to see a possible hidden structure that is obscured when working with EG alone.  Another interesting question is whether these theories could violate Lloyd's bound, analogously to how they violate the Kovtun-Son-Starinets bound on the shear viscosity to entropy density ratio $\eta/s\ge 1/(4\pi)$ \cite{Buchel:2004di,Kats:2007mq,Brigante:2007nu,Myers:2008yi,Cai:2008ph,Ge:2008ni,deBoer:2009pn,Camanho:2010ru}, which was thought to be saturated in Einstein gravity holography \cite{Kovtun:2004de}.

One of the most suitable higher curvature theories for holographic applications is Lovelock gravity \cite{Brigante:2007nu,Camanho:2009vw,Buchel:2009sk,deBoer:2009gx,Camanho:2009hu,deBoer:2009pn,Camanho:2010ru,deBoer:2011wk}, due to unique properties such as second-order equations of motion \cite{Lovelock1,Lovelock2} and the existence of a well-posed action functional \cite{Myers:1987yn,Teitelboim:1987zz}. Moreover, some of the Lovelock densities are actually predicted to appear in the effective low energy action of String Theory \cite{METSAEV1987385}, so they provide realistic corrections to the Einstein-Hilbert action.

In this paper, we compute the complexity growth of black holes in Lovelock theory using the CA proposal. While CA presents no new conceptual challenges within higher curvature gravity,  it is non-trivial to identify the correct contributions to the action coming from the null boundaries and the joints in the WDW patch.  We focus on charged black holes, since in this case the WDW patch approaches the inner and outer horizons at late time, allowing us to completely deduce the time dependent structure of the null boundary terms. On the other hand, the contribution from the joints was recently described in~\cite{CanoJoints}. Accounting for these terms, we will be able to identify an intriguing relation between the complexity growth at late times and the thermodynamic properties of the black hole.

Let us start by describing the theory and solutions of interest. The bulk action we consider is given by 
\begin{equation}\label{LoveF}
I_{\rm bulk}=\int_{\mathcal{M}}d^{d+1}x\sqrt{|g|}\left\{\mathcal{L}_{\rm grav}-\frac{1}{4g^2}F_{\mu\nu}F^{\mu\nu}\right\}\, ,
\end{equation}
where $g$ is a constant,  $F = dA$  is the Maxwell field strength, and ${\cal L}_{\rm grav}$ is the Lovelock Lagrangian~\cite{Lovelock1,Lovelock2},
\begin{align}
{\cal L}_{\rm grav}=& \frac{1}{16\pi G} \bigg[ \frac{d(d-1)}{L^2}+R 
\nonumber\\
&+ \sum_{n=2}^{\lfloor d/2\rfloor}\lambda_n \frac{(d-2n)!}{(d-2)!}(-1)^{n}L^{2n-2}\mathcal{X}_{2n} \bigg]\, ,
\end{align}
where the Euler densities $\mathcal{X}_{2n}$ are given by\footnote{The generalized Kronecker symbol is defined as $\delta^{\mu_1\mu_2\dots \mu_r}_{\nu_1\nu_2\dots\nu_r}= r!\delta^{[\mu_1}_{\nu_1}\delta^{\mu_2}_{\nu_2}\dots \delta^{\mu_r]}_{\nu_r}$. }
\begin{equation}
\mathcal{X}_{2n}= \frac{1}{2^{n}}\delta^{\mu_1\dots \mu_{2n}}_{\nu_1\dots \nu_{2n}}R^{\nu_1\nu_2}_{\mu_1\mu_2}\dots R^{\nu_{2n-1}\nu_{2n}}_{\mu_{2n-1}\mu_{2n}}\, ,
\end{equation}
and $\lambda_n$ are arbitrary dimensionless parameters.

Charged black holes in Lovelock gravity are known, see for example~\cite{Wheeler:1985qd, Cvetic:2001bk, Frassino:2014pha}, and in general the solution takes the form
\begin{eqnarray}\label{solQ}
ds^2&=&-f(r)dt^2+\frac{dr^2}{f(r)}+r^2d\Sigma_{k,d-1}^2\, ,\\
A&=&dt \frac{g}{2\sqrt{2\pi G}}\sqrt{\frac{d-1}{d-2}}\frac{q}{r^{d-2}}=\phi(r) dt\, ,
\end{eqnarray}
where $d \Sigma^2_{k , d-1}$ characterizes the constant curvature transverse geometry, with $k=+1, 0, -1$ denoting spherical, planar and hyperbolic, respectively.  The function $f$ satisfies the algebraic equation
\begin{equation}\label{BHeq}
h\left(\frac{L^2(f(r)-k)}{r^2}\right)=\frac{\omega^{d-2}L^2}{r^d}-\frac{q^2 L^2}{r^{2(d-1)}}\, ,
\end{equation}
with $h(x)$ given by the polynomial function
\begin{equation}
h(x)=1-x+\sum_{n=2}^{\lfloor d/2\rfloor}\lambda_n x^n\, .
\end{equation}
In these expressions, $q$ and $\omega$ are two integration constants that are related to the mass $M$ and the charge $Q$ of the black hole according to\footnote{We define the charge as $Q= g^{-2} \int \star F\, .$
}
\begin{eqnarray}
\omega^{d-2}&=&\frac{16\pi G M}{(d-1)\Omega_{k,d-1}}\, ,\\
q&=&\frac{g Q}{\Omega_{k,d-1}}\sqrt{\frac{8 \pi G}{(d-1)(d-2)}}\, ,
\end{eqnarray}
where $\Omega_{k,d-1}$ is the (dimensionless) volume of the transverse space. Strictly speaking, $M$ and $Q$ are the mass and the charge only in the spherically symmetric case, $k=1$. In the non-compact cases we should interpret $Q/\Omega_{k,d-1}$ as a charge density, and $M/\Omega_{k,d-1}$ as a mass density.

We require that the causal structure of the charged black holes matches figure~\ref{PenroseCharged}. While this is always the case in Einstein gravity, in Lovelock gravity the couplings must obey certain constraints in order to avoid a singularity before the inner horizon. For example, in Gauss-Bonnet gravity a sufficient condition is to demand $0 \le \lambda_2 < 1/4$, or $\lambda_3 < - (\lambda_2)^2/3$ in the cubic case.
Within this constrained class of theories, we also note that, for $d$ even, there is a special behavior that happens when the energy is smaller than 
\be\label{Mmin}
M_{\rm min} = \frac{(d-1)\Omega_{k, d-1}}{(16 \pi G)}(-k)^{d/2} L^{d-2} \lambda_{d/2}\, .
\ee
Depending on the values of the couplings, neutral solutions with $M<M_{\rm min}$ are either naked singularities --- thus, there are no black holes with mass below $M_{\rm min}$ --- or black holes with an inner horizon, as in the charged case --- see e.g.~\cite{PhysRevD.38.2434,CAI2004237,Deppe:2014oua} for details.

Let us call $r_+$ the largest root of $f$, which represents the event horizon, and $r_-$ the second root, which is the usual inner horizon of charged black holes.  The temperature of the black hole is given in terms of the derivative of $f$ as $f'(r_+)=4 \pi T_+$. On the other hand, the entropy is given by Wald's formula~\cite{Wald:1993nt}, or equivalently by the Jacobson-Myers' result~\cite{Jacobson:1993xs}--- see below~\eqref{JM} --- and it reads
\begin{align}
\label{ENT}
S_+&=\frac{r_+^{d-1}\Omega_{k,d-1}}{4 G}\left[1-\sum_{n=2}^{\lfloor d/2\rfloor}\lambda_n\left(-\frac{k L^2}{r_+^2}\right)^{n-1}\frac{n(d-1)}{(d+1-2n)}\right]\, .
\end{align}
For the sake of convenience, it will be useful to introduce as well the quantities $T_-$, $S_-$, defined analogously at $r_-$, but one should bear in mind that these do not have a natural interpretation as actual temperature and entropy.

These eternal black hole geometries should be dual to thermofield double states, created by entangling each copy of the boundary CFT as~\cite{Maldacena:2001kr},
\begin{align}
| {\rm TFD} (t_L, & t_R) \rangle = Z^{-1/2} \sum_{\alpha, \sigma} e^{- i E_{\alpha} (t_L + t_R)} 
\nonumber\\
& \times e^{-(E_{\alpha} - \mu Q_{\sigma})} |E_{\alpha}, - Q_{\sigma} \rangle | E_{\alpha}, Q_{\sigma} \rangle \, .
\end{align}

We show in figure~\ref{PenroseCharged} a schematic Penrose diagram for charged black holes in asymptotic AdS spacetimes. In our conventions, both $t_L$ and $t_R$ increase towards the upper part of the diagram, so the boost symmetry in the state is encoded in $t_R \rightarrow t_R + \Delta t$ and $t_L \rightarrow t_L - \Delta t$. Therefore, we can focus on symmetric times $t_L = t_R = t/2$ without loss of generality.

In figure \ref{PenroseCharged}, we recognize two coordinates that encode the time dependence, which we denote $r_m^{1}$ and $r_m^{2}$. We can identify how they depend on the boundary time $t$ by writing a transcendental equation involving the tortoise coordinate $r^{*}$, defined with $f(r)$ from eq. \eqref{solQ} such that
\begin{equation}
r^{*}(r) = - \int_{r}^{\infty} \frac{d r}{f(r)} \, , \qquad \lim_{r\to \infty} r^{*}(r) = 0 \, .
\end{equation}
The equations for $r_m^{1}$ and $r_m^{2}$ read
\begin{align}
&  \frac{t}{2} - r^{*}(r^1_m) = 0 \, , \quad  \frac{t}{2} + r^{*}(r^2_m) = 0 \, . 
\label{rmeeteqs}
\end{align}
As a consequence, the time derivative of these coordinates takes a simple form,
\begin{align}
& \frac{d r^1_m }{d t} = \frac{f(r^1_m)}{2} \, , \quad  \frac{d r^2_m }{d t} = - \frac{f(r^2_m)}{2} \, . 
\label{rmeetder}
\end{align}
Notice that at late times, $r_m^1$ approaches the inner horizon $r_{-}$, while $r_m^2$ approaches the outer horizon $r_{+}$.

The action calculation on the WDW patch for black holes with the causal structure depicted in figure \ref{PenroseCharged} has three possible contributions: a bulk integration, boundary contributions, and joint terms at $r_m^1$ and $r_m^2$,
\be 
I = I_{\rm bulk} + I_{\rm bdry} + I_{\rm joint} \, .
\ee 
Let us compute each of these terms.

First, we consider the contributions from the bulk action. If we express the integrand in the bulk action as a generic function of radius $r$, we have
\begin{align}
I_{\text{bulk}} &= \int_{{\rm WDW}} \!\!\!\!\! d^{d+1} x \, \sqrt{- g} \, {\cal L} = \int_{{\rm WDW}} \!\!\!\!\! dt dr I(r) \, ,
\end{align}
where the function $I(r)$ will depend on the theory of gravity under consideration.  For Lovelock gravity~\eqref{LoveF} it can be expressed as a total derivative $I(r)=\frac{d\mathcal{I}(r)}{dr}$, where
\begin{widetext}
\begin{equation}\label{Int}
\begin{aligned}
&\mathcal{I}(r)=\frac{\Omega_{k,d-1}}{16\pi G}\Bigg[-\frac{2(d-1)q^2}{r^{d-2}}+(d-1)\omega^{d-2} -r^{d-1}f'(r) \left(1-\sum_{n=2}^{\lfloor d/2\rfloor}\lambda_n\left(\frac{(f(r)-k) L^2}{r^2}\right)^{n-1}\!\!\frac{n(d-1)}{(d+1-2n)}\right)\, \Bigg]\, .
\end{aligned}
\end{equation}
\end{widetext}
The integration uses the fact that the Euler densities are total derivatives when evaluated on \eqref{solQ}~\cite{Deser:2005pc}, and the field equations~\eqref{BHeq} were used to simplify this result. 

We now repeat the calculations of~\cite{TimeDep} in the context of Lovelock theory. The bulk contribution to the action is the sum of three integrals,
\begin{align}
&I^{I}_{\text{bulk}} = 2 \int_{r_m^1}^{r_+}  I(r)  \left(\frac{t}{2} - r^{*}(r) \right) \, d r \,, \nonumber
\end{align}
\begin{align}
&I^{II}_{\text{bulk}} = 4 \int_{r_+}^{r_\text{max}}  I(r)  \left( - r^{*}(r) \right) \, d r \, , 
\\ \nonumber 
&I^{III}_{\text{bulk}} =  2 \int_{r_m^2}^{r_+}  I(r)  \left( - \frac{t}{2} - r^{*}(r) \right) \, d r  \, . 
\end{align}
Therefore, calculating the time derivative of the bulk term, using the properties of $r_m^1$ and $r_m^2$ in eqs. \eqref{rmeeteqs} and \eqref{rmeetder}, the answer depends on the differentiation with respect to time in the integrand, and the bulk contribution to the time derivative reads
\begin{equation}
\frac{d I_{\text{bulk}} }{d t} =  \int_{r^1_m}^{r^2_m} I(r) dr= \mathcal{I}(r)\bigg|_{r^1_m}^{r^2_m} \, .
\end{equation}
Then, in the late time limit $r^2_m\rightarrow r_+$, $r^1_m\rightarrow r_-$, and from \eqref{Int} we observe that this contribution takes the appealing form
\begin{equation}\label{Tbulk}
\frac{dI_{\rm bulk}}{dt}\bigg|_{t \to \infty} = \left(M - TS - \phi Q \right) \bigg|_{r_-}^{r_+} \, ,
\end{equation}
which is simply the difference of free energy one would associate to each horizon. 

Let us now consider the null boundary terms and show that their time dependence vanishes. The null boundary terms for EG were described in \cite{Parattu:2015gga, NullBoundaries}, but they are still unknown in Lovelock gravity. However, on general grounds, given a null segment $\mathcal{N}$ parametrized by $\lambda$ and with a transverse space metric $\sigma_{AB}$, the boundary contribution will have the form
\begin{equation}
I_{\rm null}=\int_{\mathcal{N}}d\lambda dx^{d-1}\sqrt{\sigma} \mathcal{Q}+\int_{\partial\mathcal{N}}dx^{d-1}\sqrt{\sigma}\mathcal{F}\, ,
\end{equation}
where, in analogy to the results in~\cite{CanoJoints}, we assume a possible contribution from the boundary of $\mathcal{N}$.  This contribution could be equivalently understood as adding a total derivative to $\mathcal{Q}$. Here $\mathcal{Q}$ and $\mathcal{F}$ are some polynomial functions of intrinsic and extrinsic curvatures and of the parameter~$\kappa$ defined as $k^\alpha \nabla_\alpha k_\beta = \kappa k_\beta$ --- see~\cite{GreatPoisson}, for the definition of the rest of the objects.


Now, let us ensure the time derivative of the null boundary term vanishes.  
Normalizing the null normal vectors on the WDW patch as for instance in \cite{TimeDep}, $r$ will be an affine parameter.  Further, since the integrands will be functions of $r$ that we denote by $Q(r)$ (for the contribution along ${\cal N}$) and $G(r)$ (for the contribution along $\partial {\cal N}$), we will have
\be 
\frac{d I_{\rm null}}{dt} \propto  Q(r_m^i) \frac{d r_m^i}{dt} + \frac{d G}{dr} \bigg|_{r=r_m^i} \frac{d r_m^i}{dt} \, .
\ee
Since these terms will be built from polynomials of intrinsic and extrinsic quantities, they will be finite (or vanishing) as $r_m^i \to r_\pm$. Since the time derivatives of $r_m^i$ vanish in this limit, we are then assured that the null boundaries make no contributions to the time derivative at late times,
\be 
\frac{d I_{\rm null}}{dt}\bigg|_{t\to\infty} = 0 \, .
\ee

Finally, we consider the contribution to the rate of change of joints at $r_m^1$ and $r_m^2$. The joint terms for intersections of null boundaries were described in~\cite{CanoJoints}, where it was found that  they are given by
\begin{equation}
I_{\rm joint}=\frac{1}{2\pi}\int_{\mathcal{C}}d\sigma a \rho_{\rm JM}\, ,
\end{equation} 
where the parameter $a$ is the same that appears in Einstein gravity~\cite{NullBoundaries}, and $\rho_{\rm JM}$ is the Jacobson-Myers entropy~\cite{Jacobson:1993xs} associated to the codimension 2 surface $\mathcal{C}$:
\begin{equation}
\label{JM}
\rho_{\rm JM}=\frac{1}{4 G}\bigg[1\!+\!\sum_{n=2}^{\lfloor d/2\rfloor}\!n\lambda_n \frac{(d-2n)!}{(d-2)!}(-1)^{n}L^{2n-2}\hat{\mathcal{X}}_{2(n-1)}\bigg]\, ,
\end{equation} 
where $\hat{\mathcal{X}}_{2(n-1)}$ is the $(n-1)$th Euler density of the induced metric.
For the case depicted in figure~\ref{PenroseCharged}, the joint contributions take the form 
\begin{equation}
I_{\rm joint} = \frac{1}{2 \pi} \left[ S (r_m^1) a (r_m^1) + S(r_m^2)  a(r_m^2) \right]\, ,
\end{equation}
where $S(r)$ evaluates to the entropy at the horizons. Following the conventions of, for instance \cite{Carmi:2016wjl, TimeDep}, 
the function $a$ at joints like those of $r_m^1$ and $r_m^2$ is given by
\begin{equation}\label{apar}
a(r) = -\log \left( \frac{|f (r)|}{\alpha^2} \right) \,, 
\end{equation}
where $\alpha$ is an arbitrary constant in the normalization of the null vector with respect to a boundary timelike vector, as described in \cite{NullBoundaries}.

The time derivative of the joint contributions to the action takes a compact and simple form at late times. For instance, at $r_m^1$ the time derivative of $S a$ takes the general form
\begin{align}\label{DerivativeOfCorner}
\frac{d \left( S \, a \right)}{d t} \bigg|_{r = r_m^1} =& \frac{1}{2} a(r_m^1) f(r_m^1) \frac{d S(r)}{d r}  \bigg|_{r = r_m^1}  
\nonumber\\
&- \frac{1}{2} S(r_m^1) f'(r_m^1) \, ,
\end{align}
where we used \eqref{rmeetder} and \eqref{apar}. The first contribution to the right hand side of eq. \eqref{DerivativeOfCorner} vanishes when $r_m^1$ approaches $r_-$ since the derivative of $S(r)$ is finite and  $\lim_{r_m^1\to r_{-}} a(r_m^1) f(r_{m}^1) = 0$. On the other hand, $f'(r_m^1)/(4\pi)$ approaches the ``temperature" $T_-$ of the horizon when evaluated at $r_-$. An analogous computation holds for $r_m^2$ approaching $r_{+}$. Then, the time derivative of the joint contribution takes the simple form
\begin{align}\label{Tjoint}
\frac{d I_{\rm joint}}{d t}\bigg|_{t \rightarrow \infty} = T S \bigg|_{r_{-}}^{r_{+}} \, .
\end{align}

Putting together the results  \eqref{Tbulk} and \eqref{Tjoint}, we have the late time complexity growth rate,
\be 
 \pi  \frac{d {\cal C}_A}{dt}\bigg|_{t\to\infty} := \pi \dot{\mathcal{C}}_A = \phi_- Q - \phi_+ Q \, ,
\ee
which holds for Lovelock theory of any order and in any dimension.

Our result can be expressed in another useful form. Introducing the dimensionless parameters $y \equiv r_-/r_+$ and $z \equiv L/r_+$ we can write,
\be 
\dot{\mathcal{C}}_A = \frac{2 M}{\pi} \left[\frac{(h_+ - y^d h_-)(1-y^{d-2})}{h_+ - y^{2(d-1)} h_-} \right]
\ee
with $h_+ := h(-k z^2)$ and $h_- := h(-k z^2/y^2)$.  We can then consider two limits of interest. First, we see that in the extremal limit $y \to 1$ and we get $\dot{\mathcal{C}}_A \to 0$. Second, we can consider the limit of vanishing charge, $y \to 0$. Here we must take note of the following result,
\be 
\lim_{y \to 0} y^d h_- = \begin{cases}
0 \quad &\text{for $d$ odd} \, ,
\\
(-k)^{d/2} \lambda_{d/2} z^d \quad &\text{for $d$ even}
\end{cases}
\ee
and so we obtain in the uncharged limit,
\be \label{neutralC}
\dot{\mathcal{C}}_A = \frac{2 (M-M_{\rm min})}{\pi} 
\ee
where it is understood that the correction to $2M$ is only present for even $d$ and $k$ non-zero\footnote{It would be interesting to investigate further the connection between topological effects and complexity as explored in \cite{Fu:2018kcp}. One possible direction would be in the context of Lovelock-Chern-Simons theories~\cite{Crisostomo:2000bb}.} --- see \eqref{Mmin} for the expression of $M_{\rm min}$. Strictly speaking, we should consider this result to hold in a regime where the mass is much larger than the charge, but the charge is still large enough that the inner horizon is not `close' to the singularity.

The vanishing of $\dot{\mathcal{C}}_A$ at extremality is in line with results from Einstein gravity~\cite{Brown:2015lvg, Brown:2015bva, NullBoundaries}. However, we note that a very interesting behavior appears in neutral black holes in even dimensions due to the appearance of $M_{\rm min}$.  As we remarked, there are two possible scenarios. The first possibility is that when $M=M_{\rm min}$ the black hole becomes zero size and then the correction in \eqref{neutralC} ensures that $\dot{\mathcal{C}}_A=0$ in that case. For $M<M_{\rm min}$ there is no black hole. The second possibility is that when $M<M_{\rm min}$ the black hole develops an internal horizon, and in that case  $\dot{\mathcal{C}}_A=0$ even if the black hole has non-vanishing temperature. Note that this same result can be obtained directly from the uncharged solution in this case, due to the two horizon causal structure, and so is true irrespective of the uncharged limit presented above. Therefore, there is a minimum mass below which black holes do not increase complexity. 
Since it is usually claimed that black holes are the fastest computers on nature,
this result would suggest that there is a minimum mass required to perform computation.

To summarize, we have carried out the first general calculation within the ``Complexity = Action'' framework taking higher curvature corrections into account.  In the late time limit, we argued that due to the WDW patch approaching  the inner and outer horizons of charged black holes that the null boundary terms are unimportant and the calculation requires only the bulk and joint terms, which are now understood~\cite{CanoJoints}. 
For spherical black holes in Gauss-Bonnet gravity, our results agree with those in \cite{Cai:2016xho}, though we note there they were computed using other methods, taking the limit of spacelike and timelike boundaries.

The complexity growth rate reduced beautifully to thermodynamic expressions,
\be \label{U+-}
\pi \dot{\mathcal{C}}_A = (F_+ + T_+S_+) - (F_- + T_-S_-) = U_+ - U_-
\ee
with $F_\pm$ the free energy associated to each horizon, and $U_\pm$ the internal energy.\footnote{Let us note that this relation was also noticed in \cite{Huang:2016fks}, where it was argued to hold in general. However, the argument relied on a number of assumptions whose validity is difficult to assess. This work was done in framework of black hole chemistry~\cite{Kastor:2009wy, Kubiznak:2016qmn}, where the mass is interpreted as the enthalpy of spacetime --- see also~\cite{Couch:2016exn}.} This result is at once surprising and suggestive. It shows that the results first obtained in~\cite{Brown:2015bva,Brown:2015lvg} are of incredibly broad scope, holding their form even in the presence of higher curvature (finite $N$) corrections.  Our calculation shows in a very transparent way the origin of this result: the bulk contribution is always the free energy and the joint contribution is always $TS$. From this, one might expect that this expression is of broader applicability than the situation considered here, and may in fact hold for any two horizon configuration. At the very least, this suggests a deep connection between the late time growth of complexity and black hole thermodynamics that merits further exploration, as for instance in \cite{Brown:2017jil}.

In principle, the prescription we have applied here will also work directly in the uncharged case. However, as noted in~\cite{Brown:2015lvg, Cai:2016xho}, there are subtleties related to the way in which one regularizes the hypersurface above the singularity --- we will discuss this further in forthcoming work~\cite{TBA}. In the charged case these problems are in general not present since the singularity is hidden behind the inner horizon. 

Lastly, let us note that while we have taken an important first step toward understanding the role of higher curvature theories in the framework of holographic complexity there remains much to explore. For example, we expect to see corrections to the complexity of formation~\cite{Chapman:2016hwi}. Furthermore, it would be interesting to explore the corrections to the full time dependence of complexity. In~\cite{TimeDep} it was found that the late time rate $2M/\pi$ is approached from above, rather than from below as Lloyd's bound would suggest. Addressing these questions would require full knowledge of the null boundary terms, which are still unknown for Lovelock gravity.

\begin{acknowledgments}
\section*{Acknowledgments}

We are pleased to thank Rob Myers and Robb Mann for useful conversations and comments. 
The work of PAC is funded by Fundaci\'on la Caixa through a ``la Caixa - Severo Ochoa" international pre-doctoral grant. The work of PAC was also supported by the MINECO/FEDER, UE grant FPA2015-66793-P and by the Spanish Research Agency (Agencia Estatal de Investigaci\'on) through the grant IFT Centro de Excelencia Severo Ochoa SEV-2016- 0597. PAC also thanks Perimeter Institute ``Visiting Graduate Fellows" program. Research at Perimeter Institute is supported by the Government of Canada through the Department of Innovation, Science and Economic Development and by the Province of Ontario through the Ministry of Research, Innovation and Science.

\end{acknowledgments}

\bibliography{mybib}{}

\end{document}